\documentclass[apjl]{emulateapj}
\usepackage{comment}
\usepackage{ifthen}

\newcommand{\ctbd}[1]{}

\newcommand{\lc}{light curve}
\newcommand{\lcs}{light curves}
\newcommand{\Lc}{Light curve}

\newcommand{\band}[1]{\ensuremath{#1}~band}

\newcommand{\chisq}{\ensuremath{\chi^2}}

\newcommand{\kms}{\ensuremath{\rm km\,s^{-1}}}
\newcommand{\ms}{\ensuremath{\rm m\,s^{-1}}}

\newcommand{\gcmc}{\ensuremath{\rm g\,cm^{-3}}}
\newcommand{\ergscmsq}{\ensuremath{\rm erg\,s^{-1}\,cm^{-2}}}

\newcommand{\vsini}{\ensuremath{v \sin{i}}}
\newcommand{\feh}{\ensuremath{\rm [Fe/H]}}

\newcommand{\vmac}{\ensuremath{v_{\rm mac}}}
\newcommand{\vmic}{\ensuremath{v_{\rm mic}}}

\newcommand{\rsun}{\ensuremath{R_\sun}}
\newcommand{\msun}{\ensuremath{M_\sun}}
\newcommand{\lsun}{\ensuremath{L_\sun}}

\newcommand{\rstar}{\ensuremath{R_\star}}
\newcommand{\mstar}{\ensuremath{M_\star}}
\newcommand{\lstar}{\ensuremath{L_\star}}

\newcommand{\teffstar}{\ensuremath{T_{\rm eff\star}}}
\newcommand{\rhostar}{\ensuremath{\rho_\star}}
\newcommand{\loggstar}{\ensuremath{\log{g_{\star}}}}

\newcommand{\mearth}{\ensuremath{M_\earth}}

\newcommand{\rpl}{\ensuremath{R_{p}}}
\newcommand{\mpl}{\ensuremath{M_{p}}}

\newcommand{\rhopl}{\ensuremath{\rho_{p}}}

\newcommand{\arstar}{\ensuremath{a/\rstar}}
\newcommand{\zrstar}{\ensuremath{\zeta/\rstar}}

\newcommand{\rjup}{\ensuremath{R_{\rm J}}}
\newcommand{\mjup}{\ensuremath{M_{\rm J}}}

\newcommand{\reffigl}[1]{Figure~\ref{fig:#1}}
\newcommand{\refsecl}[1]{\mbox{Section \ref{sec:#1}}}

\newcommand{\reftabl}[1]{Table~\ref{tab:#1}}

\newcommand{\flwof}{\mbox{FLWO 1.2\,m}}

\newcommand{\hatcurhtr}{HTR123-008}                                    
\newcommand{\hatcurfield}{123}                                         
\newcommand{\hatcurCCra}{\ensuremath{00^{\mathrm h}38^{\mathrm m}17.56{\mathrm s}}}                                  
\newcommand{\hatcurCCdec}{\ensuremath{+42{\arcdeg}27{\arcmin}47.1{\arcsec}}}                                 
\newcommand{\hatcurCCmag}{10.8}                                      
\newcommand{\hatcurCCtwomass}{2MASS~00381756+4227470}                  
\newcommand{\hatcurCCgsc}{GSC~2792-01700}                              
\newcommand{\hatcurCCtassmv}{10.812}                                   
\newcommand{\hatcurCCtwomassJmag}{\ensuremath{9.850\pm0.022}}          
\newcommand{\hatcurCCtwomassHmag}{\ensuremath{9.623\pm0.022}}          
\newcommand{\hatcurCCtwomassKmag}{\ensuremath{9.553\pm0.020}}          
\newcommand{\hatcurCCesoKmag}{\ensuremath{9.596\pm0.021}}              
\newcommand{\hatcurCCesoJKmag}{\ensuremath{0.319\pm0.032}}             
\newcommand{\hatcurLCdip}{\ensuremath{10.2}}                           
\newcommand{\hatcurLCrprstar}{\ensuremath{0.1071\pm0.0014}}            
\newcommand{\hatcurLCbsq}{\ensuremath{0.193_{-0.069}^{+0.063}}}        
\newcommand{\hatcurLCimp}{\ensuremath{0.439_{-0.098}^{+0.065}}}        
\newcommand{\hatcurLCzeta}{\ensuremath{17.73\pm0.10}}                  
\newcommand{\hatcurLCdur}{\ensuremath{0.1276\pm0.0013}}                
\newcommand{\hatcurLCdurhr}{\ensuremath{3.062\pm0.031}}                
\newcommand{\hatcurLCq}{\ensuremath{0.0460\pm0.0005}}                  
\newcommand{\hatcurLCingdur}{\ensuremath{0.0150\pm0.0014}}             
\newcommand{\hatcurLCP}{\ensuremath{2.775960\pm0.000003}}              
\newcommand{\hatcurLCPprec}{\ensuremath{2.7759602}}                    
\newcommand{\hatcurLCPshort}{\ensuremath{2.7760}}                      
\newcommand{\hatcurLCT}{\ensuremath{2455027.59293\pm0.00031}}          
\newcommand{\hatcurLCTA}{\ensuremath{2454297.5154\pm0.0009}}           
\newcommand{\hatcurLCTB}{\ensuremath{2455135.8554\pm0.0003}}           
\newcommand{\hatcurLCiblendi}{\ensuremath{0.77\pm0.08}}                
\newcommand{\hatcurSMEiteff}{\ensuremath{6175\pm80}}                   
\newcommand{\hatcurSMEizfeh}{\ensuremath{0.15\pm0.06}}                 
\newcommand{\hatcurSMEizfehshort}{\ensuremath{0.15}}                   
\newcommand{\hatcurSMEilogg}{\ensuremath{4.44\pm0.10}}                 
\newcommand{\hatcurSMEivsin}{\ensuremath{4.4\pm0.5}}                   
\newcommand{\hatcurSMEivmac}{\ensuremath{4.6}}                         
\newcommand{\hatcurSMEivmic}{\ensuremath{0.85}}                        
\newcommand{\hatcurSMEiiteff}{\ensuremath{6158\pm80}}                  
\newcommand{\hatcurSMEiizfeh}{\ensuremath{+0.17\pm0.08}}               
\newcommand{\hatcurSMEiizfehshort}{\ensuremath{+0.17}}                 
\newcommand{\hatcurSMEiilogg}{\ensuremath{4.34\pm0.06}}                
\newcommand{\hatcurSMEiivsin}{\ensuremath{3.5\pm0.5}}                  
\newcommand{\hatcurSMEiivmac}{\ensuremath{4.61}}                       
\newcommand{\hatcurSMEiivmic}{\ensuremath{0.85}}                       
\newcommand{\hatcurDSteff}{\ensuremath{6000\pm100}}                    
\newcommand{\hatcurDSlogg}{\ensuremath{4.0\pm0.25}}                    
\newcommand{\hatcurDSvsini}{\ensuremath{3.8\pm1.0}}                    
\newcommand{\hatcurDSgamma}{\ensuremath{-16.83\pm0.19}}                
\newcommand{\hatcurDSgammaNoerr}{\ensuremath{-16.83}}                  
\newcommand{\hatcurDSnumspec}{\ensuremath{7}}                          
\newcommand{\hatcurLBii}{\ensuremath{0.2166}}                          
\newcommand{\hatcurLBiii}{\ensuremath{0.3617}}                         
\newcommand{\hatcurISOm}{\ensuremath{1.22\pm0.04}}                     
\newcommand{\hatcurISOmlong}{\ensuremath{1.218\pm0.039}}               
\newcommand{\hatcurISOr}{\ensuremath{1.24\pm0.05}}                     
\newcommand{\hatcurISOrlong}{\ensuremath{1.237\pm0.054}}               
\newcommand{\hatcurISOlogg}{\ensuremath{4.34\pm0.03}}                  
\newcommand{\hatcurISOlum}{\ensuremath{1.97\pm0.22}}                   
\newcommand{\hatcurISOmv}{\ensuremath{4.03\pm0.13}}                    
\newcommand{\hatcurISOage}{\ensuremath{2.0\pm0.8}}                     
\newcommand{\hatcurISOMK}{\ensuremath{2.74\pm0.10}}                    
\newcommand{\hatcurISOJK}{\ensuremath{0.33\pm0.02}}                    
\newcommand{\hatcurISOspec}{F8}                                        
\newcommand{\hatcurRVK}{\ensuremath{531.1\pm2.8}}                      
\newcommand{\hatcurRVk}{\ensuremath{-0.030\pm0.003}}                   
\newcommand{\hatcurRVh}{\ensuremath{-0.021\pm0.006}}                   
\newcommand{\hatcurRVgammai}{\ensuremath{4.5\pm3.8}}                   
\newcommand{\hatcurRVgammaii}{\ensuremath{-434.9\pm4.2}}               
\newcommand{\hatcurRVgammaiii}{\ensuremath{-442.9\pm2.8}}              
\newcommand{\hatcurRVjitter}{\ensuremath{8.0}}                         
\newcommand{\hatcurRVfitrms}{\ensuremath{10.0}}                        
\newcommand{\hatcurRVeccen}{\ensuremath{0.036\pm0.004}}                
\newcommand{\hatcurRVomega}{\ensuremath{214\pm8}}                      
\newcommand{\hatcurPPi}{\ensuremath{86.6\pm0.7}}                       
\newcommand{\hatcurPPlogg}{\ensuremath{3.80\pm0.04}}                   
\newcommand{\hatcurPPar}{\ensuremath{7.17\pm0.28}}                     
\newcommand{\hatcurPParel}{\ensuremath{0.0413\pm0.0004}}               
\newcommand{\hatcurPPrho}{\ensuremath{2.42\pm0.35}}                    
\newcommand{\hatcurPPm}{\ensuremath{4.19\pm0.09}}                      
\newcommand{\hatcurPPmlong}{\ensuremath{4.193\pm0.094}}                
\newcommand{\hatcurPPr}{\ensuremath{1.29\pm0.07}}                      
\newcommand{\hatcurPPrlong}{\ensuremath{1.289\pm0.066}}                
\newcommand{\hatcurPPmrcorr}{\ensuremath{0.57}}                        
\newcommand{\hatcurPPteff}{\ensuremath{1626\pm40}}                     
\newcommand{\hatcurPPtheta}{\ensuremath{0.220\pm0.011}}                
\newcommand{\hatcurPPfluxavg}{\ensuremath{1.58\pm0.16}}                
\newcommand{\hatcurXsecondary}{\ensuremath{2455028.929\pm0.005}}       
\newcommand{\hatcurXsecdur}{\ensuremath{0.1234\pm0.0020}}              
\newcommand{\hatcurXsecingdur}{\ensuremath{0.0142\pm0.0013}}           
\newcommand{\hatcurPPaequiv}{\ensuremath{0.0294\pm0.0014}}             
\newcommand{\hatcurXdist}{\ensuremath{235\pm10}}                       

\newcommand{\hatcur}{HAT-P-16}
\newcommand{\hatcurb}{HAT-P-16b}

\newcommand{\hatcurSMEversion}{ii}                                       
\newcommand{\hatcurSMEteff}{\ifthenelse{\equal{\hatcurSMEversion}{i}}{\hatcurSMEiteff}{\hatcurSMEiiteff}}
\newcommand{\hatcurSMEzfeh}{\ifthenelse{\equal{\hatcurSMEversion}{i}}{\hatcurSMEizfeh}{\hatcurSMEiizfeh}}
\newcommand{\hatcurSMEzfehshort}{\ifthenelse{\equal{\hatcurSMEversion}{i}}{\hatcurSMEizfehshort}{\hatcurSMEiizfehshort}}
\newcommand{\hatcurSMElogg}{\ifthenelse{\equal{\hatcurSMEversion}{i}}{\hatcurSMEilogg}{\hatcurSMEiilogg}}
\newcommand{\hatcurSMEvsin}{\ifthenelse{\equal{\hatcurSMEversion}{i}}{\hatcurSMEivsin}{\hatcurSMEiivsin}}
\newcommand{\hatcurSMEvmac}{\ifthenelse{\equal{\hatcurSMEversion}{i}}{\hatcurSMEivmac}{\hatcurSMEiivmac}}
\newcommand{\hatcurSMEvmic}{\ifthenelse{\equal{\hatcurSMEversion}{i}}{\hatcurSMEivmic}{\hatcurSMEiivmic}}

\newboolean{emulateapj}
\setboolean{emulateapj}{false}

\newboolean{rvtablelong}
\setboolean{rvtablelong}{true}

\newboolean{astroph}
\setboolean{astroph}{false}

\shortauthors{Buchhave et al.}
\shorttitle{\hatcur\lowercase{b}}
\ifthenelse{\boolean{emulateapj}}{

}{

}

\begin{document}

\title{\hatcur\lowercase{b}: A 4\ \mjup\ planet transiting a bright star on
	an eccentric orbit
\altaffilmark{$\star$,$\star\star$}}

\author{
   L.~A.~Buchhave\altaffilmark{1,2},
   G.~\'A.~Bakos\altaffilmark{1,3},
   J.~D.~Hartman\altaffilmark{1},
   G.~Torres\altaffilmark{1},
   G.~Kov\'acs\altaffilmark{4},
   D.~W.~Latham\altaffilmark{1},
   R.~W.~Noyes\altaffilmark{1},
   G.~A.~Esquerdo\altaffilmark{1},
   M.~Everett,
   A.~W.~Howard\altaffilmark{7},
   G.~W.~Marcy\altaffilmark{7},
   D.~A.~Fischer\altaffilmark{5},
   J.~A.~Johnson\altaffilmark{6},
   J.~Andersen\altaffilmark{2,9},
   G.~F\H{u}r\'esz\altaffilmark{1},
   G.~Perumpilly\altaffilmark{1},
   D.~D.~Sasselov\altaffilmark{1},
   R.~P.~Stefanik\altaffilmark{1},
   B.~B\'eky\altaffilmark{1} 
   J.~L\'az\'ar\altaffilmark{8},
   I.~Papp\altaffilmark{8},
   P.~S\'ari\altaffilmark{8}
}
\altaffiltext{1}{Harvard-Smithsonian Center for Astrophysics, 
	Cambridge, MA}

\altaffiltext{2}{Niels Bohr Institute, Copenhagen University, Denmark}

\altaffiltext{3}{NSF Fellow}

\altaffiltext{4}{Konkoly Observatory, Budapest, Hungary}

\altaffiltext{5}{Department of Astronomy, Yale University, New Haven, CT 06511}

\altaffiltext{6}{Department of Astrophysics, California Institute of Technology,
MC 249-17, Pasadena, CA 91125, USA}

\altaffiltext{7}{Department of Astronomy, University of California,
	Berkeley, CA}

\altaffiltext{8}{Hungarian Astronomical Association, Budapest, 
	Hungary}

\altaffiltext{9}{Nordic Optical Telescope Scientific Association, 
La Palma, Canarias, Spain}

\altaffiltext{$\star$}{
Based in part on observations made with the Nordic Optical Telescope, 
operated on the island of La Palma jointly by Denmark, Finland, Iceland, 
Norway, and Sweden, in the Spanish Observatorio del Roque de los 
Muchachos of the Instituto de Astrofisica de Canarias
}

\altaffiltext{$\star\star$}{
  Based in part on observations obtained at
  the W.~M.~Keck Observatory, which is operated by the University of
  California and the California Institute of Technology. Keck time has
  been granted by NASA (N018Hr).
}
\altaffiltext{$\star\star\star$}{
All BJD values given in this paper are UTC-based (e.g. \citet{torres:2010})
}


\begin{abstract}
We report the discovery of \hatcurb{}, a transiting extrasolar planet 
orbiting the $V=\hatcurCCmag\ \rm{mag}$ \hatcurISOspec\ dwarf 
\hatcurCCgsc, with a period $P=\hatcurLCP\,\rm{d}$, transit epoch $T_c 
=\hatcurLCT$ (BJD\altaffilmark{$\star\star\star$}), and transit duration \hatcurLCdur\,d. The host star 
has a mass of \hatcurISOm\,\msun, radius of \hatcurISOr\,\rsun, 
effective temperature \hatcurSMEteff\,K, and metallicity $\feh= 
\hatcurSMEzfeh$. The planetary companion has a mass of 
\hatcurPPmlong\,\mjup, and radius of \hatcurPPrlong\,\rjup\ yielding a 
mean density of \hatcurPPrho\,\gcmc. Comparing these observed 
characteristics with recent theoretical models, we find that \hatcurb\ 
is consistent with a 1 Gyr H/He-dominated gas giant planet. \hatcurb\ 
resides in a sparsely populated region of the mass--radius diagram and 
has a non-zero eccentricity of $e=0.036$ with a significance of 
$10\sigma$.
\end{abstract}



\section{Introduction}
\label{sec:introduction}

Planets that transit their host stars play a special role in our understanding 
of the characteristics of exoplanets: their transit allows us to accurately
determine the radius and the orbital inclination of the planet from the
photometric light curve so that an actual mass can be derived from a
spectroscopic orbit of the host star. The mass and radius enable us to
infer a bulk composition of the planet, and although there are 
degeneracies associated with the bulk composition, it allows us to put
constraints on models of planetary structure and formation theories.
The incredible diversity of the over 60 discovered transiting planets,
ranging from dense planets with a higher mean density than that of
copper to strongly irradiated puffed-up planets with a mean density comparable
to that of corkwood, have baffled the
exoplanet community, and no unified theory has been established to 
explain all the systems consistently. Transiting extrasolar 
planet (TEP) discoveries are primarily the result of dedicated ground--based
searches, such as SuperWASP \citep{pollacco:2006}, HATNet 
\citep{bakos:2004}, TrES \citep{alonso:2004} and XO \citep{pmcc:2005}, 
and space-borne searches, such as CoRoT \citep{baglin:2006} and Kepler 
\citep{borucki:2010}.

Since its commissioning in 2003, the Hungarian-made Automated Telescope 
Network \citep[HATNet;][]{bakos:2004} survey has been one of the major 
contributors to the discoveries of TEPs. HATNet has discovered over a 
dozen TEPs since 2006 by surveying bright stars ($8\lesssim I \lesssim 
12.5$\,mag) in the Northern hemisphere and has now covered approximately 
11\% of the Northern sky. HATNet consists of six wide field automated 
telescopes; four of these are located at the Fred Lawrence Whipple 
Observatory (FLWO) in Arizona, and two on the roof of the Submillimeter 
Array hangar (SMA) of the Smithsonian Astrophysical Observatory (SAO) in Hawaii. 
In this paper we report a new TEP discovery of HATNet, called \hatcurb.

The structure of the paper is as follows. In \refsecl{obs} we summarize
the observations, including the photometric detection, and follow-up
observations. In \refsecl{analysis} we describe the analysis of the
data, such as the stellar parameter determination
(\refsecl{stelparam}), ruling out blend scenarios (\refsecl{blend}), and
global modeling of the data (\refsecl{globmod}). We discuss our
findings in \refsecl{discussion}.

\section{Observations}
\label{sec:obs}
\subsection{Photometric detection}
\label{sec:detection}
\begin{figure}[!ht]
\plotone{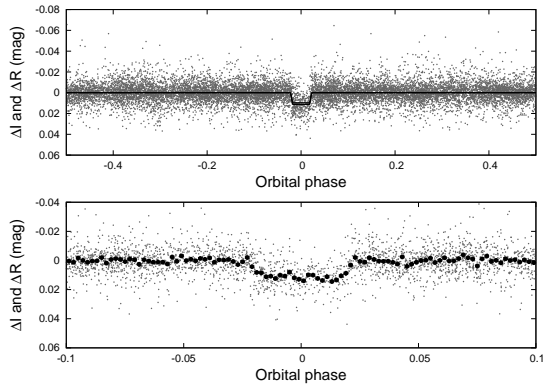}
\caption{
	Unbinned \lc{} of \hatcur{} including all 12,552 instrumental
    \band{I} and \band{R} 5.5 minute cadence measurements obtained with
    the HAT-6, HAT-7, HAT-8 and HAT-9 telescopes (see the
    text for details), and folded with the period of $P=
    \hatcurLCPprec$\,days (which is the result of the fit described in
    \refsecl{analysis}). The \band{I} and \band{R} data are merged in 
    this figure. The solid line shows the transit model fit to the 
    light curve (\refsecl{globmod}). In the lower panel, the large solid circles 
    represent the binned average of the photometric measurements
    with a bin size of 0.005 days.
\label{fig:hatnet}}
\end{figure}

The transits of \hatcurb{} were detected with the HAT-6 and HAT-7 
telescopes in Arizona and the HAT-8 and HAT-9 telescopes in Hawaii. The 
star \hatcurCCgsc{} lies in the intersection of 3 separate HATNet fields 
internally labeled as \hatcurfield, 163 and 164. Field \hatcurfield\ was 
observed with an $I$ filter and a 2K$\times$2K CCD, while fields 163 and 164 
were observed through $R$ filters with 4K$\times$4K CCDs. All three 
fields were observed with a 5 minute exposure time and at a 5.5 minute 
cadence in the period between July 2007 and September 2008 during which over  
12,000 exposures were gathered for the three fields. Each image contains between 27,000 and 
76,000 stars down to a magnitude of $I \sim 13$ for field 123 and $R 
\sim 15$ for fields 163 and 164, yielding a photometric precision for 
the brightest stars in the field of $\sim 2$\,mmag for field 123 and 
$\sim 5$\,mmag for fields 163 and 164.

Frame calibration, astrometry and aperture photometry were done in an
identical way to recent HATNet TEP discoveries, as described in
\cite{bakos:2009} and \cite{pal:2009}. The resulting \lcs\ were decorrelated (cleaned of
trends) using the External Parameter Decorrelation technique in
``constant'' mode \citep[EPD; see][]{bakos:2009} and the Trend
Filtering Algorithm \citep[TFA; see][]{kovacs:2005}. The \lcs{} were
searched for periodic box-like signals using the Box Least-Squares
method \citep[BLS; see][]{kovacs:2002}. A significant signal was
detected in the \lc{} of \hatcurCCgsc{} (also known as
\hatcurCCtwomass{}; $\alpha=\hatcurCCra$, $\delta=\hatcurCCdec$;
J2000; $V=\hatcurCCtassmv$; \citealp{droege:2006}), with a depth of
$\sim\hatcurLCdip$\,mmag, and a period of $P=\hatcurLCPshort$\,days. 
The dip had a relative duration (first to last contact) of
$q\approx\hatcurLCq$, corresponding to a total duration of
$Pq\approx\hatcurLCdurhr$~hr (see \reffigl{hatnet}).

\subsection{Reconnaissance Spectroscopy}
\label{sec:recspec}
Transiting planet candidates found by ground-based, wide-field 
photometric surveys must undergo a rigorous vetting process to eliminate 
the many astrophysical systems mimicking transiting planets (called 
false positives), the rate of which has proved to be much higher than 
the occurrence of true planets (10 to 20 times higher). Low signal to 
noise ratio (S/N) high-resolution reconnaissance spectra are used to extract 
stellar parameters such as effective temperature, gravity, metallicity 
and rotational and radial velocities to rule out these false positives. 
Examples of false positives which are discarded are eclipsing binaries and
triple systems. The latter can be either hierarchical or chance 
alignment systems where the light of the eclipsing pair of stars is 
diluted by the light of a third brighter star. Rapidly rotating and/or 
hot host stars whose spectrum is unsuitable for high precision velocity 
work are also discarded.

We acquired \hatcurDSnumspec\ reconnaissance spectra with the CfA
Digital Speedometer \citep[DS;][]{latham:1992} mounted on the FLWO
1.5\,m Tillinghast Reflector between December 2008 and January 2009.
The extracted modest precision radial velocities gave a mean absolute
RV=\hatcurDSgammaNoerr\,\kms\, with an rms of $0.51\ \kms$, which is consistent with no detectable RV
variation. The stellar parameters derived from the spectra 
\citep{torres:2002}, including the effective
temperature $\teffstar=\hatcurDSteff\,K$, surface gravity $\loggstar
=\hatcurDSlogg$ (log cgs) and projected rotational velocity $\vsini=
\hatcurDSvsini\,\kms$, correspond to a \hatcurISOspec\ dwarf.

\subsection{High resolution, high S/N spectroscopy}
\label{sec:hispec}

The high significance of the transit detection by HATNet, together with the stellar
spectral type and small RV variations encouraged us to gather
high--resolution, high S/N spectra to determine the orbit of the system.
We have taken 21 spectra between August and October 2009 using the
FIbre--fed \'Echelle Spectrograph (FIES) at the 2.5\,m Nordic Optical
Telescope (NOT) at La Palma, Spain \citep{djupvik:2009}. We used the medium and the
high--resolution fibers ($1\farcs3$ projected diameter) with 
resolving powers of $\lambda/\Delta\lambda \approx 46,\!000$ and
$67,\!000$, respectively, giving a wavelength coverage of $\sim
3600-7400\ \rm{\AA}$. Recently, FIES has also been used to obtain a
spectroscopic orbit for one of the first Kepler planets, namely
Kepler-7b \citep{latham:2010}.

We also used the HIRES instrument \citep{vogt:1994} on the Keck~I
telescope located on Mauna Kea, Hawaii. We obtained 6 exposures with an
iodine cell, plus 1 iodine-free template with Keck-I/HIRES.  The
observations were made on 6 nights between 2009 July 3 and 2009 October
29. The width of the spectrometer slit used on HIRES was $0\farcs86$,
resulting in a resolving power of $\lambda/\Delta\lambda \approx
55,\!000$, with a wavelength coverage of $\sim3800-8000$\,\AA\@. The
iodine gas absorption cell was used to superimpose a dense forest of
$\mathrm{I}_2$ lines on the stellar spectrum and establish an accurate
wavelength fiducial \citep[see][]{marcy:1992}. Relative RVs in the
Solar System barycentric frame were derived as described by
\cite{butler:1996}, incorporating full modeling of the spatial and
temporal variations of the instrumental profile.

\begin{figure} [ht]
\plotone{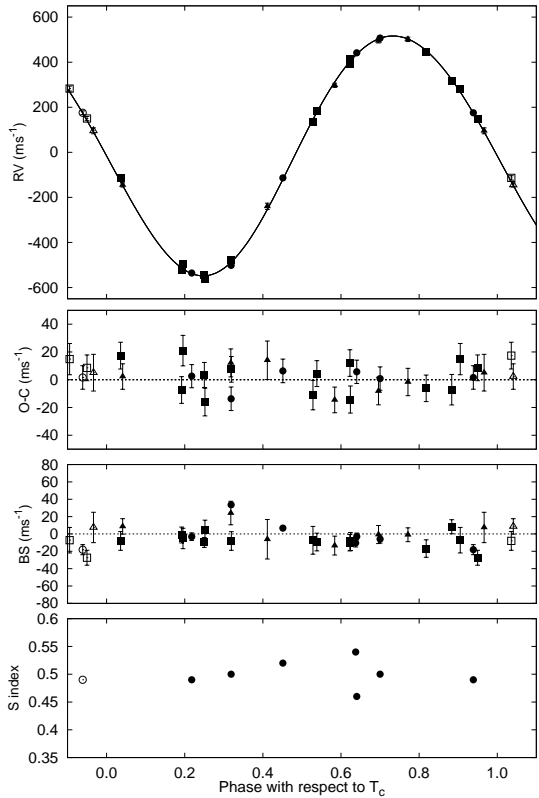}
\caption{
    Top: RV measurements from the Nordic Optical Telescope and  
    Keck for \hatcur{}, along with an orbital
    fit, shown as a function of orbital phase, using our best-fit
    period (see \refsecl{analysis}). 
    Velocities from NOT using the medium and high resolution fiber are shown 
    as squares and triangles, respectively, and the Keck velocities are shown 
    as circles. Zero phase is defined by the
    transit midpoint. The center-of-mass velocity has been subtracted. 
    Note that the error bars include the stellar jitter, added in
    quadrature to the formal errors given by the spectrum reduction
    pipeline.
	Second panel: phased residuals after subtracting the orbital fit
	(also see \refsecl{analysis}). The rms variation of the residuals
	is about $\hatcurRVfitrms$\,\ms.
	Third panel: bisector span variations (BS) including the template spectrum
    (\refsecl{blend}).
	Bottom: relative S activity index values which have only been calculated 
   for the Keck spectra. 
    Note the different vertical scales of the panels.
\label{fig:rvbis}}
\end{figure}

The final RV data and their errors (for both instruments) are listed in 
\reftabl{rvs}. The folded data are plotted in \reffigl{rvbis}. The 
systemic gamma velocities (that are the result of the global modeling, 
as laid out in \refsecl{analysis}) have been subtracted to ensure a common 
zero--point. The best orbital fit (see \refsecl{analysis}) is superimposed in 
the figure.

Since this is the first time the NOT has been used to determine the
orbit of a HATNet planet, we describe briefly the extraction of the
radial velocities.
We have built a custom pipeline to rectify and cross correlate spectra
from \'Echelle spectrographs with the goal of providing very precise
radial velocities. The first step in the extraction process is to
remove the bias level and crop the raw images. To effectively remove
cosmic rays, the observation is divided into three separate exposures,
which enables us to combine the raw images using median filtering,
removing virtually all cosmic rays. We use a flat field to trace the
\'Echelle orders and we rectify the spectra by using the ``optimal
extraction'' algorithm \citep{hewett:1985,horne:1986}. The blaze
function is determined by fitting cubic splines to a combined high S/N
flat field exposure and is saved separately in order to preserve the original
flux in the stellar exposure. By dividing the normalized blaze function
into the rectified flat field spectrum, we can determine the pixel to
pixel variations and correct for these. The scattered light in the
two--dimensional raw image is determined and removed by masking out the
signal in the \'Echelle orders and fitting the inter--order scattered
light flux with a two--dimensional polynomial.

Thorium argon (ThAr) calibration images are obtained through the
science fiber before and after the stellar observation. The two
calibration images are combined to form the basis for the fiducial
wavelength calibration. We have determined that the best wavelength
solution is achieved by choosing an exposure time that saturates the
strong argon lines but enhances the forest of weaker thorium lines.

FIES is not a vacuum spectrograph and is only temperature controlled to 
$0.1\,^{\circ}\rm{C}$. Consequently, the radial velocity errors are 
dominated by our limited ability to correct for shifts due to pressure, 
humidity and temperature variations. In order to successfully remove 
these large variations ($> 1.5\,\kms$), it is critical that the ThAr 
light travels through the same light path as the stellar light and thus 
acts as an effective proxy to remove these variations. We have therefore 
chosen to interleave the stellar observation between two ThAr exposures 
instead of using the simultaneous ThAr technique, which may not exactly 
describe the induced shifts in the science fiber. The centers of the 
ThAr lines are found by fitting a Gaussian function to the line profiles 
and a two-- dimensional fifth order Legendre polynomial is used to 
describe the wavelength solution.

Once the spectra have been extracted, a multi--order cross correlation
is performed order by order. First, the spectra are interpolated to a
common oversampled log wavelength scale with the same monotonic
wavelength increment in all orders. A high and low pass filter is
applied in the Fourier domain and the ends of the spectra are apodized
with a cosine bell function. The orders are cross correlated using a
Fast Fourier Transform (FFT) and the cross correlation function (CCF)
for each order is co--added. This automatically weights each order by
its flux, giving more weight to the orders with more photons. This CCF
peak is fitted with a Gaussian function to determine its center. The
internal precision is estimated by finding the radial velocity for each
order and the RV error is thus $\sigma=RMS(v)/\sqrt{N}$, where $v$ is
the RV of the individual orders and $N$ is the number of orders. For
the final radial velocities, a template spectrum is constructed by
shifting and co--adding all the observed spectra, and the individual
spectra are cross correlated against this co--added template spectrum
to minimize the contribution of noise in the template.

\ifthenelse{\boolean{emulateapj}}{
    \begin{deluxetable*}{lrrrrrr}
}{
    \begin{deluxetable}{lrrrrrr}
}
\tablewidth{0pc}
\tablecaption{
	Relative radial velocity and bisector span variation
	measurements of \hatcur{}.
	\label{tab:rvs}
}
\tablehead{
	\colhead{BJD} & 
	\colhead{RV\tablenotemark{a}} & 
	\colhead{\ensuremath{\sigma_{\rm RV}}} & 
	\colhead{BS} & 
	\colhead{\ensuremath{\sigma_{\rm BS}}} & 
	\colhead{Inst.} \\
	\colhead{\hbox{~~~~(2,455,000$+$)~~~~}} & 
	\colhead{(\ms)} & 
	\colhead{(\ms)} &
	\colhead{(\ms)} &
   \colhead{(\ms)} &
	\colhead{} &
}
\startdata
\ifthenelse{\boolean{rvtablelong}}{
$ 048.64476 $ \dotfill & $  -137.7 $ & $      4.6 $ & $    -13.4 $ & $    10.9  $ & $    \rm{FIES\ h}$\\
$ 049.70638 $ \dotfill & $  -339.5 $ & $     10.5 $ & $      7.5 $ & $    17.5  $ & $    \rm{FIES\ h}$\\
$ 050.68239 $ \dotfill & $  -911.9 $ & $      6.1 $ & $     24.1 $ & $    13.6  $ & $    \rm{FIES\ h}$\\
$ 051.73142 $ \dotfill & $    60.3 $ & $      5.8 $ & $     -0.7 $ & $    10.4  $ & $    \rm{FIES\ h}$\\
$ 052.68956 $ \dotfill & $  -579.8 $ & $      4.4 $ & $      8.7 $ & $     8.9  $ & $    \rm{FIES\ h}$\\
$ 053.71751 $ \dotfill & $  -674.7 $ & $     11.3 $ & $     -6.1 $ & $    22.8  $ & $    \rm{FIES\ h}$\\
$ 054.71713 $ \dotfill & $    66.2 $ & $      5.7 $ & $     -0.9 $ & $     8.0  $ & $    \rm{FIES\ h}$\\

$ 107.58955 $ \dotfill & $     0.7 $ & $      5.2 $ & $    -16.8 $ & $    10.1  $ & $    \rm{FIES\ m}$\\
$ 108.62791 $ \dotfill & $  -965.2 $ & $      5.3 $ & $     -1.2 $ & $     9.2  $ & $    \rm{FIES\ m}$\\
$ 109.56094 $ \dotfill & $  -310.4 $ & $      7.0 $ & $     -7.3 $ & $    16.0  $ & $    \rm{FIES\ m}$\\
$ 110.54715 $ \dotfill & $  -128.9 $ & $      7.5 $ & $      8.3 $ & $     8.1  $ & $    \rm{FIES\ m}$\\
$ 111.56509 $ \dotfill & $  -989.4 $ & $      4.1 $ & $     -9.8 $ & $     5.7  $ & $    \rm{FIES\ m}$\\
$ 112.60304 $ \dotfill & $   -53.7 $ & $      5.5 $ & $     -8.7 $ & $    10.4  $ & $    \rm{FIES\ m}$\\
$ 113.50725 $ \dotfill & $  -294.7 $ & $      4.7 $ & $    -27.4 $ & $     8.6  $ & $    \rm{FIES\ m}$\\
$ 114.53303 $ \dotfill & $  -924.1 $ & $      4.5 $ & $     -8.1 $ & $    10.7  $ & $    \rm{FIES\ m}$\\
$ 116.52227 $ \dotfill & $  -558.0 $ & $      5.3 $ & $     -8.0 $ & $    10.8  $ & $    \rm{FIES\ m}$\\
$ 122.51752 $ \dotfill & $  -941.1 $ & $      7.5 $ & $     -5.2 $ & $    11.8  $ & $    \rm{FIES\ m}$\\
$ 123.46870 $ \dotfill & $  -262.4 $ & $      4.7 $ & $     -9.3 $ & $    10.2  $ & $    \rm{FIES\ m}$\\
$ 124.48749 $ \dotfill & $  -161.9 $ & $      7.9 $ & $     -7.2 $ & $    14.7  $ & $    \rm{FIES\ m}$\\
$ 125.44928 $ \dotfill & $  1008.8 $ & $      5.9 $ & $      4.0 $ & $    11.9  $ & $    \rm{FIES\ m}$\\
$ 126.47958 $ \dotfill & $   -29.8 $ & $      5.2 $ & $    -10.1 $ & $     9.4  $ & $    \rm{FIES\ m}$\\

$ 017.09285 $ \dotfill & $  -531.7 $ & $      2.3 $ & $     -3.1 $ & $     4.3  $ & $    \rm{HIRES}$\\
$ 019.09413 $ \dotfill & $   179.5 $ & $      2.6 $ & $    -18.1 $ & $     5.8  $ & $    \rm{HIRES}$\\
$ 107.08869 $ \dotfill & \nodata  & \nodata  & $    -10.3 $ & $     4.6  $ & $    \rm{HIRES}$\\
$ 107.09645 $ \dotfill & $   445.3 $ & $      2.4 $ & $     -2.9 $ & $     3.9  $ & $    \rm{HIRES}$\\
$ 108.97959 $ \dotfill & $  -497.6 $ & $      2.6 $ & $     33.5 $ & $     1.7  $ & $    \rm{HIRES}$\\
$ 112.12355 $ \dotfill & $  -109.3 $ & $      2.8 $ & $      6.7 $ & $     2.1  $ & $    \rm{HIRES}$\\
$ 135.02181 $ \dotfill & $   510.9 $ & $      2.5 $ & $     -5.8 $ & $     4.6  $ & $    \rm{HIRES}$\\
}{
$ 048.64476 $ \dotfill & $  -137.7 $ & $      4.6 $ & $    -13.4 $ & $    10.9  $ & $    \rm{FIES\ h}$\\
$ 049.70638 $ \dotfill & $  -339.5 $ & $     10.5 $ & $      7.5 $ & $    17.5  $ & $    \rm{FIES\ h}$\\
$ 050.68239 $ \dotfill & $  -911.9 $ & $      6.1 $ & $     24.1 $ & $    13.6  $ & $    \rm{FIES\ h}$\\
}
\enddata
\tablenotetext{a}{
	The fitted zero-point that is on an arbitrary scale (denoted as
	$\gamma_{\rm rel}$ in \refsecl{globmod}) has {\em not} been
	subtracted from the velocities.
}
\ifthenelse{\boolean{rvtablelong}}{
	\tablecomments{
		For the iodine-free template exposure we do not
		measure the RV but do measure the BS. Such template exposures
		can be distinguished by the missing RV value. The Inst.  column
		refers to the instrument used, i.e.~the FIES spectrograph at
		the NOT using the medium and high resolution fibers or the
		HIRES spectrograph at Keck I. \ensuremath{\sigma_{\rm RV}} and 
		\ensuremath{\sigma_{\rm BS}} are formal statistical errors.
	}
}{
    \tablecomments{
		For the iodine-free template exposure we do not
		measure the RV but do measure the BS. Such template exposures
		can be distinguished by the missing RV value. The Inst.  column
		refers to the instrument used, i.e.~the FIES spectrograph at
		the NOT using the medium and high resolution fibers or the
		HIRES spectrograph at Keck I. \ensuremath{\sigma_{\rm RV}} and 
		\ensuremath{\sigma_{\rm BS}} are formal statistical errors.
	}
} 
\ifthenelse{\boolean{emulateapj}}{
    \end{deluxetable*}
}{
    \end{deluxetable}
}

\subsection{Photometric follow-up observations}
\label{sec:phot}

\begin{figure}[!ht]
\plotone{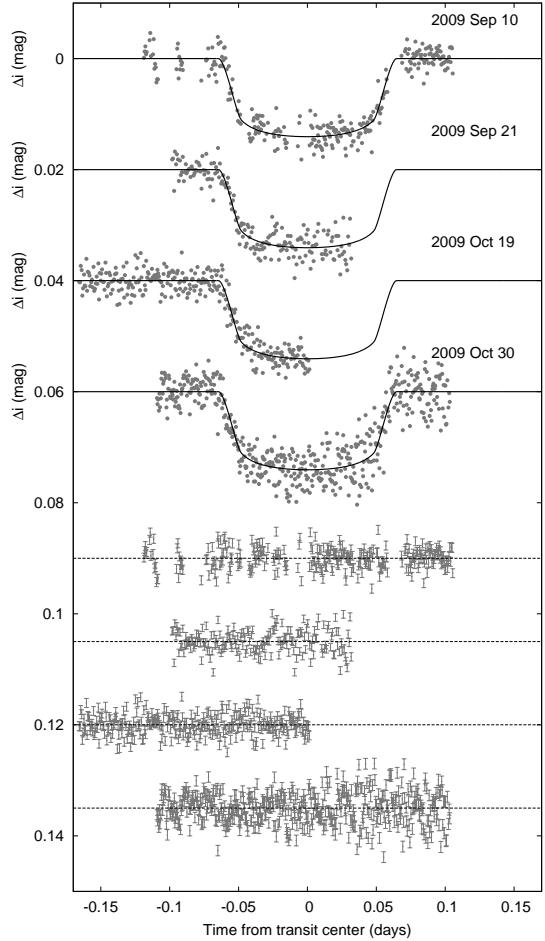}
\caption{
	Unbinned instrumental \band{i} transit \lcs{}, acquired with
    KeplerCam at the \flwof{} telescope.
	Superimposed are the best-fit transit model \lcs.  At the bottom of
	the figure we show the residuals from the fit.  Error bars
	represent the photon and background shot noise, plus the readout
	noise.
\label{fig:lc}}
\end{figure}
To confirm the transit signal and obtain high-precision light curves for 
modeling the system, we conducted photometric follow-up observations 
with the KeplerCam CCD on the \flwof{} telescope. We observed 4 transit 
events of \hatcur{} on the nights of 2009 September 10, September 21, 
October 19 and October 30 (\reffigl{lc}). On the four nights, $303, 181, 
293$ and $471$ frames were acquired with a cadence of $41, 52, 41$ and 
$32$ seconds ($30, 40, 30$ and $20$ seconds of exposure time) in the 
Sloan $i$ band, respectively.

The reduction of the images, including frame calibration, astrometry,
and photometry were performed as described in \citet{bakos:2009}. We
also performed EPD and TFA to remove trends simultaneously with the
light curve modeling (for more details, see \refsecl{analysis}). The
final \lcs{} are shown in the upper plots of \reffigl{lc}, superimposed
with our best-fit transit \lc{} models (see also \refsecl{analysis});
the photometry is provided in \reftabl{phfu}.

\begin{deluxetable}{lrrrr}
\tablewidth{0pc}
\tablecaption{Follow-up photometry of \hatcur\label{tab:phfu}}
\tablehead{
	\colhead{BJD} & 
	\colhead{Mag\tablenotemark{a}} & 
	\colhead{\ensuremath{\sigma_{\rm Mag}}} &
	\colhead{Mag(orig)\tablenotemark{b}} & 
	\colhead{Filter} \\
	\colhead{\hbox{~~~~(2,400,000$+$)~~~~}} & 
	\colhead{} & 
	\colhead{} &
	\colhead{} & 
	\colhead{}
}
\startdata
$ 55085.69759 $ & $   0.00368 $ & $   0.00082 $ & $   9.93719 $ & $ i$\\
$ 55085.69808 $ & $   0.00111 $ & $   0.00082 $ & $   9.93426 $ & $ i$\\
$ 55085.69860 $ & $  -0.00132 $ & $   0.00082 $ & $   9.93260 $ & $ i$\\
$ 55085.69911 $ & $  -0.00200 $ & $   0.00082 $ & $   9.93017 $ & $ i$\\
$ 55085.69962 $ & $   0.00022 $ & $   0.00082 $ & $   9.93280 $ & $ i$\\
\enddata
\tablenotetext{a}{
	The out-of-transit level has been subtracted. These magnitudes have
	been subjected to the EPD and TFA procedures, carried out
	simultaneously with the transit fit.
}
\tablenotetext{b}{
	These are raw magnitude values without application of the EPD
	and TFA procedures.
}
\tablecomments{
    This table is available in a machine-readable form in the
    online journal. A portion is shown here for guidance regarding
    its form and content.
}
\end{deluxetable}

\section{Analysis}
\label{sec:analysis}
\subsection{Properties of the parent star}
\label{sec:stelparam}
The stellar atmospheric parameters were derived from the template
spectrum obtained with the Keck/HIRES instrument. We used the
Spectroscopy Made Easy (SME) analysis package of \cite{valenti:1996},
along with the atomic-line database of \cite{valenti:2005}. This
yielded the following {\em initial} values and uncertainties (which we
have conservatively increased to include our estimates of the
systematic errors):
effective temperature $\teffstar=\hatcurSMEiteff$\,K, stellar surface
gravity $\loggstar=\hatcurSMEilogg$\,(cgs), metallicity
$\feh=\hatcurSMEizfeh$\,dex, and projected rotational velocity
$\vsini=\hatcurSMEivsin\,\kms$.

We could now use the effective temperature and the surface gravity found 
by SME to determine other stellar
parameters, such as the mass, \mstar, and radius, \rstar, using model
isochrones. However, there may be strong correlations between effective
temperature, gravity and metallicity in the spectroscopic determination
of these parameters. Also, the effect of \loggstar\ on the spectral line
shapes is typically subtle, and as a result \loggstar\ is generally a
rather poor luminosity indicator. Instead, we used the 
\arstar, the normalized semimajor axis (related to \rhostar, the mean
stellar density), which can be derived directly
from the \lcs\ \citep{sozzetti:2007}.

The stellar parameters were determined simultaneously with the modeling
of the \lcs\ and radial velocities, as described next. We began by
adopting the values of \teffstar, \feh, and \loggstar\ from the SME
analysis to fix the quadratic limb-darkening coefficients from the
tabulations by \cite{claret:2004}, which are needed to model the \lcs\
($a_{i}$, $b_{i}$). This modeling yields the probability distribution
of \arstar\ via a Monte Carlo approach, which is described fully in
\refsecl{globmod}. We then used the \arstar\ distribution together with
Gaussian distributions for \teffstar\ and \feh, with 1$\sigma$
uncertainties as reported previously, to estimate \mstar\ and \rstar\
by comparison with the stellar evolution models of \citet{yi:2001}. 
This was done for each of 15,000 simulations. Parameter combinations
corresponding to unphysical locations in the H-R diagram (4\% of the
trials) were ignored, and replaced with another randomly drawn
parameter set. The resulting stellar parameters and their uncertainties
were determined from the {\em a posteriori} distributions obtained in
this way.

In particular, the resulting surface gravity of $\loggstar=
\hatcurISOlogg$ is somewhat different from that derived in the initial
SME analysis, which is not surprising in view of the possible strong
correlations among \teffstar, \feh, and the surface gravity. Therefore, in a
second iteration we adopted this value of \loggstar\ and held it fixed
in a new SME analysis, adjusting only \teffstar, \feh, and \vsini.
This gave 
$\teffstar=\hatcurSMEiiteff$\,K, 
$\feh=\hatcurSMEiizfeh$, and 
$\vsini=\hatcurSMEiivsin$\,\kms, 
which we adopt as final atmospheric values for the star. Finally, we
repeated the calculation of the stellar mass and radius with these
values and the stellar evolution models, and obtained
\mstar=\hatcurISOmlong\,\msun, \rstar=\hatcurISOrlong\,\rsun\ and
\lstar=\hatcurISOlum\,\lsun, along with other parameters summarized in
\reftabl{stellar}.

\begin{figure}[!ht]
\plotone{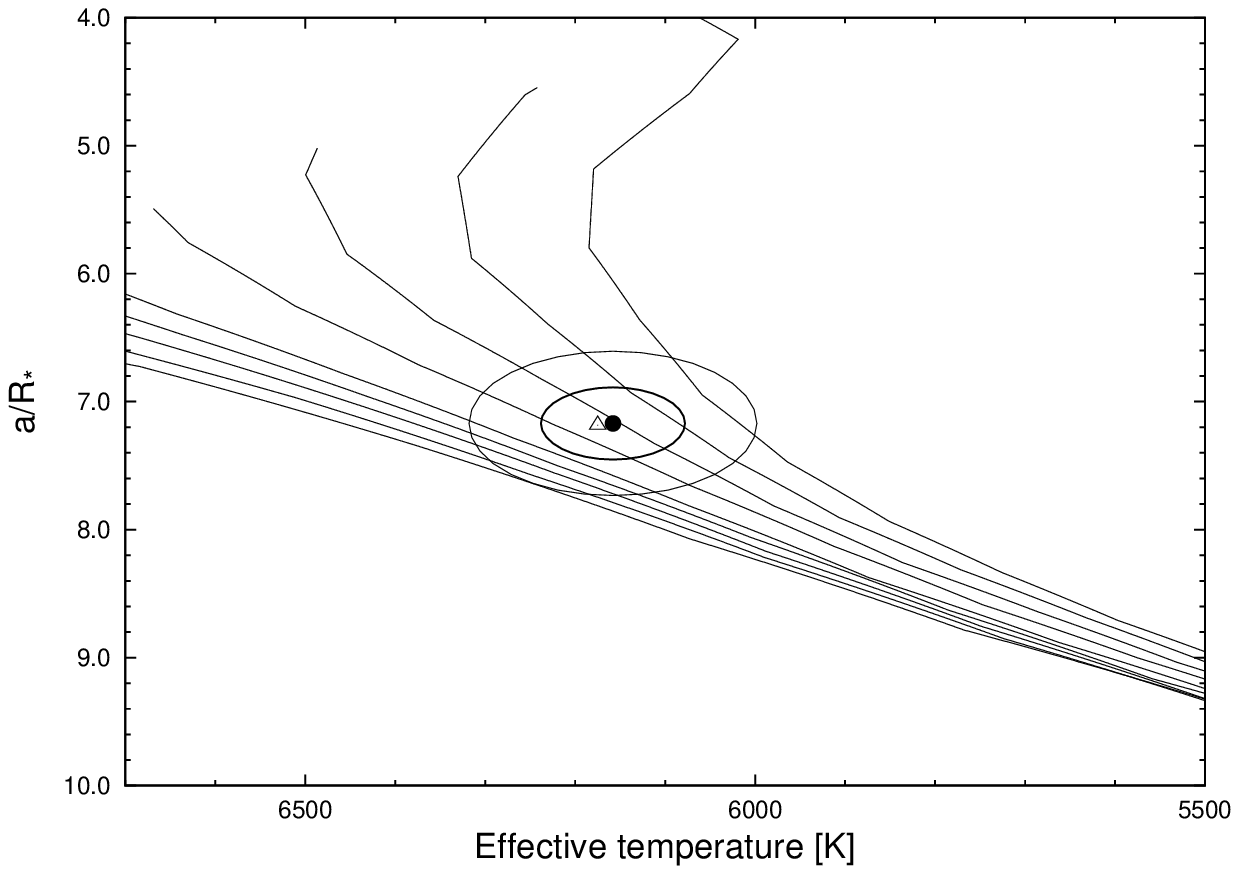}
\caption{
	Stellar isochrones from \citet{yi:2001} for metallicity
        \feh=\hatcurSMEiizfehshort\ and ages 0.2, 0.4, 0.6, 0.8,
        1.0, 1.5, 2.0, 2.5 and 3.0\,Gyr (left to right).  The final choice of
        $\teffstar$ and \arstar\ are marked and encircled by the
        1$\sigma$ and 2$\sigma$ confidence ellipsoids. The initial
        value of \teffstar\ and \arstar\ from the first SME analysis
        and subsequent \lc\ analysis is marked with a triangle that is
		barely offset from the final value (indicated by a filled
        circle).
\label{fig:iso}}
\end{figure}

The model isochrones from \citet{yi:2001} for metallicity
\feh=\hatcurSMEzfehshort\ are plotted in \reffigl{iso}, with the final
choice of effective temperature $\teffstar$ and \arstar\ marked, and
encircled by the 1$\sigma$ and 2$\sigma$ confidence ellipsoids.

The stellar evolution modeling provides color indices that may be
compared against the measured values as a sanity check. The best
available measurements are the near-infrared magnitudes from the 2MASS
Catalogue \citep{skrutskie:2006},
$J_{\rm 2MASS}=\hatcurCCtwomassJmag$, 
$H_{\rm 2MASS}=\hatcurCCtwomassHmag$ and 
$K_{\rm 2MASS}=\hatcurCCtwomassKmag$;
which we have converted to the photometric system of the models
(ESO) using the transformations by \cite{carpenter:2001}. The
resulting measured color index is $J-K=\hatcurCCesoJKmag$.
This is within 1$\sigma$ of the predicted value from the isochrones
of $J-K=\hatcurISOJK$. The distance to the object may be computed
from the absolute $K$ magnitude from the models ($M_{\rm
K}=\hatcurISOMK$) and the 2MASS $K_s$ magnitude, which has the
advantage of being less affected by extinction than optical magnitudes.
The result is $\hatcurXdist$\,pc, where the uncertainty excludes
possible systematics in the model isochrones that are difficult to
quantify.

\newcommand{\hatcurisoshort}{YY}
\newcommand{\hatcurlumind}{\arstar}
\begin{deluxetable}{lll}
\tablewidth{0pc}
\tablecaption{
	Stellar parameters for \hatcur{}
	\label{tab:stellar}
}
\tablehead{
	\colhead{Parameter}	&
	\colhead{Value} &
	\colhead{Source}
}
\startdata
$\teffstar$ (K)\dotfill         &  \hatcurSMEteff   & SME\tablenotemark{a}\\
$\feh$ (dex)\dotfill            &  \hatcurSMEzfeh   & SME                 \\
$\vsini$ (\kms)\dotfill         &  \hatcurSMEvsin   & SME                 \\
$\vmac$ (\kms)\dotfill          &  \hatcurSMEvmac   & SME                 \\
$\vmic$ (\kms)\dotfill          &  \hatcurSMEvmic   & SME                 \\
$\gamma_{\rm RV}$ (\kms)\dotfill         &  \hatcurDSgamma       & DS      \\
$a_{i}$\dotfill			&  \hatcurLBii		& SME+Claret\tablenotemark{b}  \\
$b_{i}$\dotfill			&  \hatcurLBiii		& SME+Claret          \\
$\mstar$ ($\msun$)\dotfill      &  \hatcurISOmlong   & \hatcurisoshort+\hatcurlumind+SME\tablenotemark{c}\\
$\rstar$ ($\rsun$)\dotfill      &  \hatcurISOrlong   & \hatcurisoshort+\hatcurlumind+SME         \\
$\loggstar$ (cgs)\dotfill       &  \hatcurISOlogg    & \hatcurisoshort+\hatcurlumind+SME         \\
$\lstar$ ($\lsun$)\dotfill      &  \hatcurISOlum     & \hatcurisoshort+\hatcurlumind+SME         \\
$V$ (mag)\dotfill				&  \hatcurCCtassmv   & TASS                                      \\
$M_V$ (mag)\dotfill             &  \hatcurISOmv      & \hatcurisoshort+\hatcurlumind+SME         \\
$K$ (mag,ESO)                   &  \hatcurCCesoKmag  & 2MASS+Carpenter\tablenotemark{d}          \\
$M_K$ (mag,ESO)\dotfill         &  \hatcurISOMK      & \hatcurisoshort+\hatcurlumind+SME         \\
Age (Gyr)\dotfill               &  \hatcurISOage     & \hatcurisoshort+\hatcurlumind+SME         \\
Distance (pc)\dotfill           &  \hatcurXdist      & \hatcurisoshort+\hatcurlumind+SME\\
\enddata
\tablenotetext{a}{
	SME = ``Spectroscopy Made Easy'' package for analysis of
	high-resolution spectra \cite{valenti:1996}. These parameters
	depend primarily on SME, with a small dependence on the iterative
	analysis incorporating the isochrone search and global modeling of
	the data, as described in the text.
}
\tablenotetext{b}{
	SME+Claret = Based on the SME analysis and tables of quadratic
    limb-darkening coefficients from \citet{claret:2004}.
}
\tablenotetext{c}{
	\hatcurisoshort+\hatcurlumind+SME = YY2 isochrones
	\citep{yi:2001}, \arstar\ luminosity indicator, and SME results.
}
\tablenotetext{d}{Based on the relations from \citet{carpenter:2001}.}
\end{deluxetable}

\subsection{Excluding blend scenarios}
\label{sec:blend}

We performed a line bisector analysis of the observed spectra following 
\cite{torres:2007} to explore the possibility that the main cause of the 
radial velocity variations is actually distortions in the spectral line 
profiles due to stellar activity or a nearby unresolved eclipsing 
binary. The bisector analysis was performed on the Keck spectra as described
in \S~5 of \cite{bakos:2007a}; the spectra from the NOT were analyzed in a 
similar manner.

The resulting bisector span variations are plotted in \reffigl{rvbis}
and have a low amplitude compared to the orbital semi--amplitude and
show no correlation with the radial velocities. We therefore conclude
that the velocity variations are real, and that the star is transited by
a close-in giant planet.

\subsection{Global modeling of the data}
\label{sec:globmod}

Our model for the follow-up \lcs\ used analytic formulae based on
\citet{mandel:2002} for the eclipse of a star by a planet, where the
stellar flux is described by quadratic limb-darkening. The limb
darkening coefficients are based on the SME results
(\refsecl{stelparam}) and the tables provided by \citet{claret:2004}
for the $i$ band. The transit shape was parametrized by the normalized
planetary radius $\rpl/\rstar$, the square of the impact
parameter $b^2$, and the reciprocal of the half mid-ingress to mid-egress
duration of the transit
$\zrstar$. We chose these parameters because of their simple geometric
meanings and the fact that they show negligible correlations
\citep[see][]{bakos:2009}. Our model for the HATNet data was the
simplified ``P1P3'' version of the \citet{mandel:2002} analytic
functions (an expansion by Legendre polynomials), for the reasons
described in \citet{bakos:2009}.
Following the formalism presented by \citet{pal:2009}, the RV curve was
modeled as an eccentric Keplerian orbit with semiamplitude $K$,
and Lagrangian orbital elements $(k,h)=e\times(\cos\omega,\sin\omega)$

We assumed that there is a strict periodicity in the individual transit
times. In practice, we assigned the transit number $N_{tr}=0$ to the
first high quality follow-up \lc\ gathered on 2009 September 10.  The
adjusted parameters in the fit were the first transit center observed
with HATNet, $T_{c,-284}$, and the last high quality transit center
observed with the \flwof\ telescope, $T_{c,+18}$. The transit center
times for the intermediate transits were interpolated using these two
epochs and the $N_{tr}$ transit number of the actual event
\citep{bakos:2007b}. The model for the RV data contains the ephemeris
information through the $T_{c,-284}$ and $T_{c,+18}$ variables. 
Altogether, the 8 parameters describing the physical model were 
$T_{c,-284}$, $T_{c,+18}$, $\rpl/\rstar$, $b^2$, $\zrstar$, $K$, 
$k=e\cos\omega$, $h=e\sin\omega$. Nine additional parameters were
related to the instrumental configuration. These are the three
instrumental blend factors $B_{\rm inst,i}$ of HATNet (one for each
of the three fields), which account for possible dilution of the transit in the
HATNet \lc\ due to the wide (20\arcsec\ wide FWHM) PSF and possible
crowding, the three HATNet out-of-transit magnitudes, $M_{\rm
0,HN,i}$, and three relative RV zero--points $\gamma_{\rm rel,j}$ (one
each for the Keck, high-resolution FIES, and medium-resolution FIES
observations).

We extended our physical model with an instrumental model that
describes the systematic variations of the data. This was done in a
similar fashion to the analysis presented in \citet{bakos:2009}. The
HATNet photometry has already been EPD- and TFA-corrected before the
global modeling, so we only considered systematic correction of the
follow-up \lcs. We chose the ``ELTG'' method, i.e.~EPD was performed in
``local'' mode with EPD coefficients defined for each night, and TFA
was performed in ``global'' mode using the same set of stars and TFA
coefficients for all nights. The underlying physical model was based on
the \citet{mandel:2002} analytic formulae, as described earlier.  The
five EPD parameters were the hour angle (characterizing a monotonic
trend that linearly changes over time), the square of the hour angle,
and the stellar profile parameters (equivalent to FWHM, elongation,
position angle).
The functional form of the above parameters contained six
coefficients, including the auxiliary out-of-transit magnitude of the
individual events. The EPD parameters were independent for all four
nights, implying 24 additional coefficients in the global fit. For the
global TFA analysis we chose 20 template stars that had good quality
measurements for all nights and on all frames, implying an additional
20 parameters in the fit. Thus, the total number of fitted parameters
was 17 (physical model) + 24 (local EPD) + 20 (global TFA) = 61.

The joint fit was performed as described in \citet{bakos:2009}.  
We minimized \chisq\ in the parameter space by using a hybrid
algorithm, combining the downhill simplex method \citep[AMOEBA;
see][]{press:1992} with the classical linear least squares algorithm. 
Uncertainties for the parameters were derived using the Markov Chain
Monte-Carlo method \citep[MCMC, see][]{ford:2006,pal:2009}.

The eccentricity of the system appeared as significantly non-zero
($k=\hatcurRVk$, $h=\hatcurRVh$).
The best-fit results for the relevant physical parameters are
summarized in \reftabl{planetparam}. We also list
the RV ``jitter,'' which is a component of assumed astrophysical noise
intrinsic to the star that we add in quadrature to the RV measurement
uncertainties in order to have $\chi^{2}/{\rm dof}=1$ from the RV
data for the global fit. In addition, some auxiliary parameters (not
listed in the table) are:
$T_{\mathrm{c},-284}=\hatcurLCTA$~(BJD),
$T_{\mathrm{c},+18}=\hatcurLCTB$~(BJD),
$\gamma_{\rm rel}=\hatcurRVgammai$\,\ms, 
$\gamma_{\rm rel,FIEShi}=\hatcurRVgammaii$\,\ms (FIES, high
resolution),
$\gamma_{\rm rel,FIESmed}=\hatcurRVgammaiii$\,\ms (FIES, medium
resolution), $B_{instr,123}=\hatcurLCiblendi$,
$B_{instr,163}=0.93\pm0.04$, 
$B_{instr,164}=0.96\pm0.04$.
Note that these gamma velocities do \emph{not} correspond to the true center 
of mass RV of the system, but are only relative offsets. The true systemic  
velocity of the system is RV=\hatcurDSgamma\,\kms found by cross--correlating 
the observed spectrum against a library template spectrum.
The planetary parameters and their uncertainties can be derived by the
direct combination of the {\em a posteriori} distributions of the \lc,
RV and stellar parameters.  We found that the planet is fairly massive
with $\mpl=\hatcurPPmlong\,\mjup$, and compact with radius of
$\rpl=\hatcurPPrlong\,\rjup$, yielding a mean density
of $\rho_p=\hatcurPPrho$\,\gcmc. The final planetary parameters are
summarized at the bottom of Table~\ref{tab:planetparam}.

\begin{deluxetable}{lr}
\tablewidth{0pc}
\tablecaption{Orbital and planetary parameters\label{tab:planetparam}}
\tablehead{
	\colhead{~~~~~~~~~~~~~~~Parameter~~~~~~~~~~~~~~~} &
	\colhead{Value}
}
\startdata
\sidehead{\Lc{} parameters}
~~~$P$ (days)             \dotfill    & $\hatcurLCP$              \\
~~~$T_c$ (${\rm BJD}$)    \dotfill    & $\hatcurLCT$              \\
~~~$T_{14}$ (days)
      \tablenotemark{a}   \dotfill    & $\hatcurLCdur$            \\
~~~$T_{12}=T_{34}$ (days)
    \tablenotemark{a}     \dotfill    & $\hatcurLCingdur$         \\
~~~$\arstar$              \dotfill    & $\hatcurPPar$             \\
~~~$\zrstar$              \dotfill    & $\hatcurLCzeta$           \\
~~~$\rpl/\rstar$          \dotfill    & $\hatcurLCrprstar$        \\
~~~$b^2$                  \dotfill    & $\hatcurLCbsq$            \\
~~~$b \equiv a \cos i/\rstar$
                          \dotfill    & $\hatcurLCimp$            \\
~~~$i$ (deg)              \dotfill    & $\hatcurPPi$ \phn         \\

\sidehead{RV parameters}
~~~$K$ (\ms)              \dotfill    & $\hatcurRVK$              \\
~~~$k_{\rm RV}$\tablenotemark{b} 
                          \dotfill    & $\hatcurRVk$              \\
~~~$h_{\rm RV}$\tablenotemark{b}
                          \dotfill    & $\hatcurRVh$              \\
~~~$e$                    \dotfill    & $\hatcurRVeccen$          \\
~~~$\omega$                  \dotfill    & $\hatcurRVomega^\circ$   \\
~~~RV jitter (\ms)           \dotfill    & \hatcurRVjitter           \\
~~~RV rms from fit (\ms)           \dotfill    & \hatcurRVfitrms           \\

\sidehead{Secondary eclipse parameters}
~~~$T_s$ (BJD)            \dotfill    & $\hatcurXsecondary$       \\
~~~$T_{s,14}$             \dotfill    & $\hatcurXsecdur$          \\
~~~$T_{s,12}$             \dotfill    & $\hatcurXsecingdur$       \\

\sidehead{Planetary parameters}
~~~$\mpl$ ($\mjup$)       \dotfill    & $\hatcurPPmlong$          \\
~~~$\rpl$ ($\rjup$)       \dotfill    & $\hatcurPPrlong$          \\
~~~$C(\mpl,\rpl)$
    \tablenotemark{c}     \dotfill    & $\hatcurPPmrcorr$         \\
~~~$\rhopl$ (\gcmc)       \dotfill    & $\hatcurPPrho$            \\
~~~$a$ (AU)               \dotfill    & $\hatcurPParel$           \\
~~~$\log g_p$ (cgs)       \dotfill    & $\hatcurPPlogg$           \\
~~~$T_{\rm eq}$ (K)       \dotfill    & $\hatcurPPteff$           \\
~~~$\Theta$\tablenotemark{d}               \dotfill    & $\hatcurPPtheta$          \\
~~~$\langle F \rangle$ ($10^{9}$\ergscmsq) \tablenotemark{e}
                          \dotfill    & $\hatcurPPfluxavg$        \\
\enddata
\tablenotetext{a}{
	\ensuremath{T_{14}}: total transit duration, time
	between first to last contact;
	\ensuremath{T_{12}=T_{34}}: ingress/egress time, time between first
	and second, or third and fourth contact.
}
\tablenotetext{b}{
    Lagrangian orbital parameters derived from the global modeling, 
    and primarily determined by the RV data. 
}
\tablenotetext{c}{
	Correlation coefficient between the planetary mass \mpl\ and radius
	\rpl.
}
\tablenotetext{d}{
	The Safronov number is given by $\Theta=\frac{1}{2}(V_{\rm
	esc}/V_{\rm orb})^2=(a/\rpl)(\mpl / \mstar )$
	\citep[see][]{hansen:2007}.
}
\tablenotetext{e}{
	Incoming flux per unit surface area.
}
\end{deluxetable}



\section{Discussion}
\label{sec:discussion}

We present the discovery of \hatcurb\ with a period of $P=
\hatcurLCPshort\ \rm{d}$, a mass of $\mpl=\hatcurPPm\ \mjup$ and a
radius of $\rpl=\hatcurPPr\ \rjup$. Currently, there are only a
handful of planets residing near \hatcurb\ in the
mass--radius diagram. CoRoT-2b \citep{alonso:2008} and CoRoT-6b
\citep{fridlund:2010} have similar masses and periods, but both have
eccentricities consistent with zero. HD80606 \citep{naef:2001} and
HD17156 \citep{fischer:2007} are also similar in mass, but both
planets have long periods and very eccentric orbits. WASP-10b
($P=3.09$\,d, $e=0.06, \mpl=2.96\ \mjup, \rpl=1.08\ \rjup$, see
\citealt{christian:2009,johnson:2009}) is probably the transiting
planet which most resembles \hatcurb\, with a similar period and
eccentricity, albeit slightly lower mass and radius. At $V=
\hatcurCCmag\ \rm{mag}$, \hatcur\ is one of the brightest stars known to host a 
$\sim 4\ \mjup$ planet.

\begin{figure*}[!ht]
\plotone{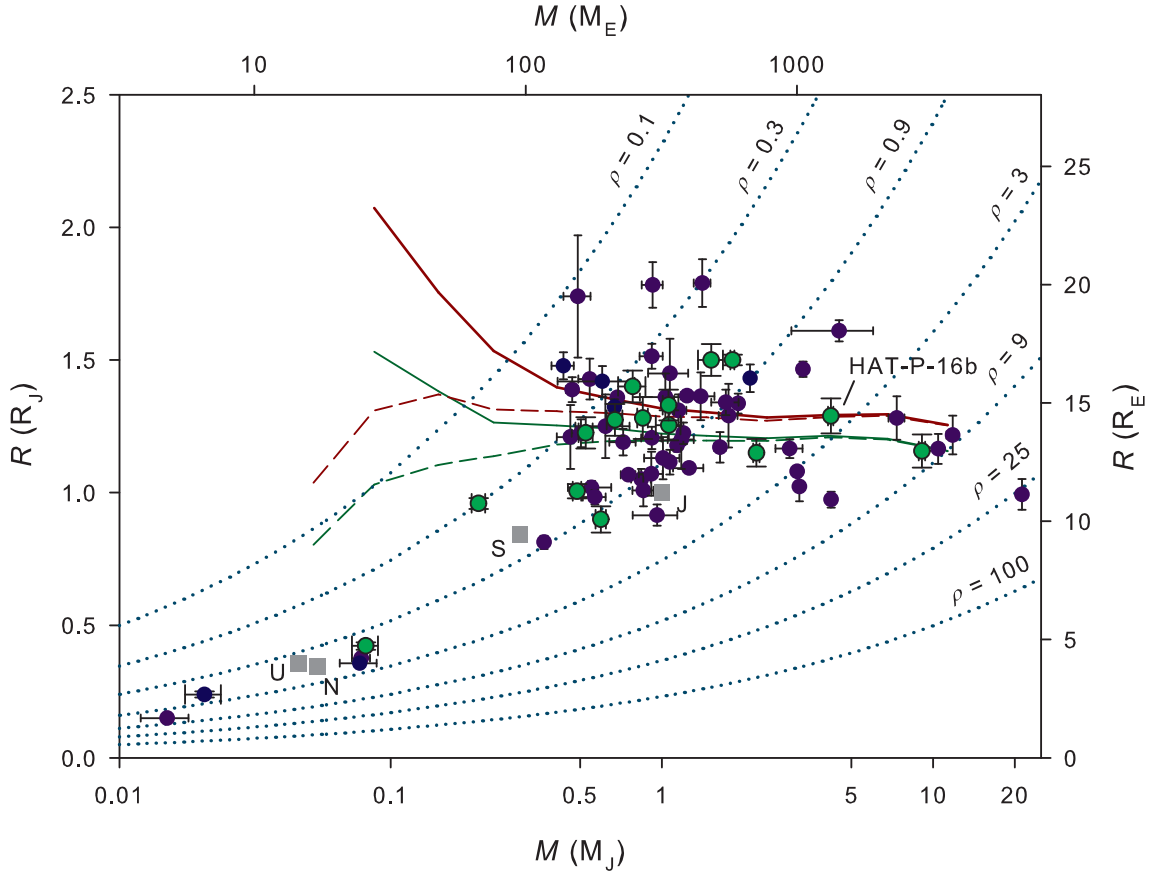}
\caption{
   Mass--radius diagram of currently known TEPs. HATNet planets are
   show in green and \hatcurb\ is labeled explicitly. The Solar system
   planets are shown as filled gray squares. Isodensity curves (in $\rm{g\ cm^{-3}}$) are
   plotted as dotted lines. Overlaid are planetary 1.0 Gyr (brown, 
   the upper set of  lines) and
   4.5 Gyr (green, the lower set of lines) isochrones from \citet{fortney:2007}
   for H/He-dominated
   planets with core mases of $M_{c}=0\ \mearth$ (solid) and 
   $M_{c}=10\ \mearth$ (dashed), interpolated to the solar equivalent 
   semimajor axis of \hatcurb.
\label{fig:exomr}}
\end{figure*}

We have compared \hatcurb\ to the theoretical models from
\citet{fortney:2007} by interpolating the models to the
solar-equivalent semi--major axis of $a_{equiv}=\hatcurPPaequiv\
\rm{AU}$, the result of which can be seen overplotted in
\reffigl{exomr}. We find that the mass and radius of \hatcurb\ are quite
consistent with the 1 Gyr model and conclude that \hatcurb\ is most
likely a H/He-dominated planet.

The radial velocity dataset used to characterize \hatcurb\ consist of 
three different velocity sets from two different telescopes (FIES medium 
and high resolution spectra from the NOT and HIRES spectra from Keck-I), 
requiring additional free parameters in the global fit to account for 
the arbitrary velocity offset between the measurements. Nevertheless, 
the different observations yield very similar semi-amplitude and we find 
an rms from the best fit model of only $\hatcurRVfitrms$\,\ms.  The rich 
dataset consisting of 27 RV observations allows us to accurately 
determine the small non-zero eccentricity of the orbit as $e=0.036$ with 
a high significance of $10\sigma$, with $k \neq 0$ at $10\sigma$ and $h 
\neq 0$ at $3\sigma$. The eccentricity is consistent and clearly evident 
even when fitting separate orbits to the three different velocity 
datasets.

Most short--period TEPs have eccentricities consistent with zero, as is 
expected because of circularization driven by tidal dissipation within 
the planet with a typical timescale substantially less than 1 Gyr (see 
\citealt{jackson:2008, mardling:2007}). In fact, the estimated 
circularization timescale for \hatcurb\ is $\tau_{cir} \sim 0.03\ 
\rm{Gyr}$ (when the tidal dissipation
parameter is assumed to be $Q_p = 10^5$), which is much less than the estimated $\sim 2\ \rm{Gyr}$ age 
of the system. On the other hand, \citet{matsumura:2008} argue that  
circularization timescales could be significantly higher and the 
planets with eccentric orbits may simply be in the process of being 
circularized. \citet{adams:2006} argue that for multiple systems with a 
hot Jupiter as the inner planet, secular excitation of the eccentricity 
by companion planets could explain the non--zero eccentricity of these 
systems. Thus the origin of the eccentricity of \hatcurb\ remains 
unclear. 

With a stellar rotation of \hatcurSMEiivsin\ \kms and an impact 
parameter of $b \sim 0.43$, \hatcur\ is a favorable target for measuring 
the Rossiter--McLaughlin (RM) effect. This effect can be used to determine the
alignment of the planet's orbital angular momentum vector with the stellar
spin axis \citep{winn:2009}.  The estimated amplitude of the RM effect is 
$47\ \ms$ for \hatcurb.

\acknowledgements 

HATNet operations have been funded by NASA grants NNG04GN74G, NNX08AF23G 
and SAO IR\&D grants. Work of G.\'A.B.~and J.~Johnson were supported by 
the Postdoctoral Fellowship of the NSF Astronomy and Astrophysics 
Program (AST-0702843 and AST-0702821, respectively). GT acknowledges 
partial support from NASA grant NNX09AF59G. We acknowledge partial 
support also from the Kepler Mission under NASA Cooperative Agreement 
NCC2-1390 (D.W.L., PI). G.K.~thanks the Hungarian Scientific Research 
Foundation (OTKA) for support through grant K-81373. Based in part on 
observations made with the Nordic Optical Telescope, operated on the 
island of La Palma jointly by Denmark, Finland, Iceland, Norway, and 
Sweden, in the Spanish Observatorio del Roque de los Muchachos of the 
Instituto de Astrofisica de Canarias.  This research has made use of 
Keck telescope time granted through NASA (N018Hr).



\end{document}